MASS LOSS ON THE RED GIANT BRANCH: PLASMOID-DRIVEN WINDS ABOVE THE RGB BUMP

D. J. Mullan and J. MacDonald

Dept. of Physics and Astronomy, University of Delaware, Newark DE 19716

ABSTRACT

The onset of cool massive winds in evolved giants is correlated with an evolutionary feature on the red giant branch (RGB) known as the "bump". Also at the bump, shear instability in the star leads to magnetic fields that occur preferentially on small length-scales. Pneuman (1983) has suggested that the emergence of small-scale flux tubes in the Sun can give rise to enhanced acceleration of the solar wind as a result of plasmoid acceleration **(the so-called "melon-seed mechanism"). In this paper, we examine Pneuman's formalism to determine if it may shed some light on the process that drives mass-loss in stars above the RGB bump. Because we do not currently have detailed information for some of the relevant physical parameters, we are not yet able to derive a detailed model: instead, our goal in this paper is to explore a "proof-of-concept". Using parameters that are known to be plausible in cool giants, we find that the total mass loss rate from such stars can be replicated. Moreover, we find that the radial profile of the wind speed in such stars can be steep or shallow depending on the fraction of the mass-loss rate which is contained in the plasmoids: this is consistent with empirical data which indicate** that the velocity profiles of winds from cool giants span a range of steepnesses.

Key words: stars: mass loss – stars: evolution – stars: magnetic field – Sun: solar wind -- MHD

 1. INTRODUCTION

Physical processes of various kinds have been identified as being responsible for mass loss from stars in different regions of the Hertzsprung-Russell (HR) diagram. In solar-like stars, the finite pressure of the coronal gas leads to hydrodynamic expansion that successfully replicates many of the properties of the "slow" solar wind emerging from the Sun at low latitudes (Parker 1963). Also in the Sun, the faster wind that is observed to emerge from coronal holes indicates that an extra acceleration arises due to the deposit of additional momentum and energy in the supersonic wind. The additional momentum has been attributed to outward streaming waves (Jacques 1977). Alternatively, in an attempt to model time-steady mass loss over a broad range of the HR diagram, dissipation of turbulence due to counter-streaming Alfven waves has been modelled by Cranmer & Saar 2011[CS]: the stars to which CS attempt to fit their steady-state model have mass loss rates spanning a range of 12 orders of magnitude.

Among the cool giants of spectral types from K to mid-M, despite the occurrence of mass loss rates that exceed the solar value by as much as 6 orders of magnitude, Harper (2018) points out that the mechanisms for driving such mass loss rates are "very poorly understood. The winds cannot rely mainly on radiation, or on variability-generated acoustic waves or shocks…Some form of magnetic activity…is likely to be the cause of these ubiquitous mass outflows….a quantitative physical model for mass loss is lacking for these stars."

We note that CS "emphasize that a complete description of late-type stellar winds requires the incorporation of other physical processes besides Alfven waves and turbulence." Among the additional processes that may be relevant, CS list episodic flares or coronal mass ejections, large-amplitude pulsations, and radiative driving. Of these processes, pulsations and radiative forces are predominant only in certain regions of the HR diagram, e.g. in supergiants.

Our interest in the present paper is focused on the first item in the CS list, i.e. "episodic flares or coronal mass ejections". Such phenomena are expected to lead to measurable variability in the winds. Evidence for variability in mass loss from cool giants was indeed reported in the profiles of the emission cores of the MgII h and k lines (Mullan & Stencel 1982). Quantitatively, the optical depth of certain winds from cool evolved stars has been observed to change by factors of 2 - 6, while the terminal speeds change by > 20 km s$^{-1}$ (Mullan et al. 1998). Moreover, clumps of hot chromospheric plasma have been identified in the wind of certain supergiants, although the cooler plasma (with temperature of 1000 - 3000 K) dominates the mass of the wind (Harper & Brown 2006).

An important aspect of the physical process(es) responsible for accelerating a stellar wind may become accessible **if observers can successfully extract information about the velocity profile of the wind as a function of radial** **location. Such a velocity** profile indicates where energy is being imparted to the wind: **specifically, if the acceleration behavior close to the star can be extracted, the information would be** "particularly important" (Crowley et al. 2009). Observations of wind lines in eclipsing binaries such as the ζ Aurigae systems are useful in this regard, but velocity profiles have also been obtained for some isolated giants (see Fig. 1 of Crowley et al. 2009). The data suggest a variety of velocity profiles in various stars. For example, acceleration is rapid close to the surface of an isolated cool giant (Arcturus, where terminal speed is reached at an altitude of only $0.8R_*$ above the surface). But in the case of binaries, the onset of acceleration is in some cases much more gradual (having reached only 0.7 times terminal speed at a radial distance of 7 $R_*$), while in other cases, there is a postponement of onset of acceleration (out to ~3 $R_*$) followed by onset of rapid acceleration.

In the present paper, **rather than considering the broad sweep of the mass loss process which was adopted by Cranmer and Saar (2011), we focus on mass loss from stars which occupy a specific position** in the HR diagram, namely, the red giant branch (RGB). The reason for our interest in the RGB is that Crowley et al. (2009) have pointed out the following: "Despite much observational and theoretical work, the mass loss process from cool giant and supergiant stars (*in particular on the first red giant branch- FRGB*) is still poorly understood" (our emphasis added). In the present paper, we focus not only on the RGB as a whole, but even more specifically on stars which have evolved through a feature known as the "RGB bump".

The plan of the paper is as follows. In Section 2, we summarize the total mass loss that occurs on the RGB, and describe the evolutionary feature known as the bump. In Section 3, we recapitulate the shear instability results of MacDonald & Mullan (2003). In Section 4, we summarize some of the previous models that have been proposed for driving mass loss from cool giant stars. **The principal goal of the present work is to examine the work of Pneuman (1983): this is summarized in Section 5, where we apply it quantitatively to the problem of mass loss from cool giants.** (Note that, based on an earlier paper by Pneuman [1968], Mullan [1980] had made an attempt to model mass loss from cool giants, but the discussion in Mullan's 1980 paper was almost entirely qualitative.) Discussion and conclusions are given in Section 6.

## 2. THE (FIRST) RED GIANT BRANCH (FRGB)

When a low-mass star evolves away from the main sequence, it moves up along the red giant branch (RGB) for the first time. We will not be interested here in what happens on the second (asymptotic) giant branch, where pulsational instability and radiation-driven dust processes may dominate the mass loss rate (Höfner & Olofsson 2018). In what follows, we shall use the abbreviation RGB to refer to the *first* red giant branch. In order to drive mass loss from stars on the RGB, we must rely on physical processes that do *not* require pulsational instability or radiative driving. In this paper, we propose a mass-loss mechanism that satisfies this condition.

### 2.1. Integrated mass loss on the (first) RGB

In order to obtain at least an order of magnitude estimate of the mass loss rate on the RGB after a star has passed through the bump, we need **to know the values of** two numbers: (i) the total amount of mass lost on the RGB, and (ii) the time-scale for a star to evolve from the bump to the tip of the RGB. As regards (i), McDonald et al. (2011) have derived the integrated mass loss in two globular clusters 47 Tuc and omega Cen: the results are 0.22 $M_\odot$ and (0.2-0.25) $M_\odot$. As regards (ii), Bharat Kumar et al. (2015) cite time-scales for low-mass stars ranging in mass from 1 to 2 $M_\odot$: the time-scales range from $10^8$ yr (for 1 $M_\odot$) to $2\times10^7$ yr (for 2 $M_\odot$). Using these numbers, we see that the average mass loss rate on RGB may be as small as $2\times10^{-9}$ $M_\odot$ yr$^{-1}$, or as large as $10^{-8}$ $M_\odot$ yr$^{-1}$.

For comparison, we note that in an analysis of a particularly interesting subset of 7 giants (namely, those which are not only Li-rich K giants but which also exhibit IR excess), Bharat Kumar et al. (2015) have derived mass loss rates ranging from $4\times10^{-9}$ $M_\odot$ yr$^{-1}$ to $2\times10^{-7}$ $M_\odot$ yr$^{-1}$. These empirical mass loss rates overlap with the above estimates of RGB mass loss rates, although the empirical rates for the subset of 7 giants span a wider range. For present purposes, we will adopt a mass loss rate of $10^{-8}$ $M_\odot$ yr$^{-1}$ that is consistent with both ranges of mass loss rates. With this choice, **and converting to c.g.s. units for future reference, we find that** the mass loss rate on the RGB is of order $10^{18}$ g s$^{-1}$.

To be sure, there is no guarantee that the mass loss rate will have a constant value at all points of the RGB: in fact, it seems likely that near the tip of the RGB, where gravity is weaker, mass loss may be more copious, reaching values of a few times $10^{-8}$ $M_\odot$ yr$^{-1}$ (Mauas et al. 2006). But in the present paper, where we examine a **proof-of-concept** for one particular model of magnetically driven mass loss, we wish to determine if a consistent model of the wind from a RGB star can be derived which yields a mass loss rate of order $10^{-8}$ $M_\odot$ yr$^{-1}$.

Such a mass loss rate exceeds by as much as 3 orders of magnitude the values predicted by the widely used "Reimers law" (as modified by Schröder & Cuntz 2005) for stars located at the RGB bump (see Bharat Kumar et al. 2015, their Table 3). Thus, although the (modified) "Reimers law" may offer reasonably reliable estimates for mass loss rates over a broad region in the HR diagram, the empirical mass loss rates which occur locally on the RGB (and in particular above the bump) are found to be considerably larger than the (modified) Reimers law **would** predict.

This leads us to suggest that an effective physical mechanism for mass loss may be operative above the RGB bump that is not available over broad regions of the HR diagram.

### 2.2. The "bump": why does it occur?

To understand the origin of the "bump", we show in Fig. 1 how different regions inside a star of mass $M_* = 1.3$ $M_\odot$ change their **location and** extent as a function of time. **In Figure 1, time is listed** in units of billions of years (b.y.) relative to a **certain** zero-point when the star was homogeneous and on the main sequence. (**The zero-point of time is labelled** at the top left of the figure by the letter A.) **In order to illustrate various features which occur in the course of evolution, the time axis is split into 3 panels. Regions which are plotted with dark and dotted stripes in the left-hand panel and also in the middle panel indicate regions** where H → He burning is generating energy at a rate in excess of 10 erg g$^{-1}$ s$^{-1}$; regions with dots only are locations where H-burning is occurring at a smaller rate, although still consuming hydrogen and creating a local gradient in molecular weight. Letter B near the top left (at age 2.25 b.y.) denotes when the **(small)** convective core ceases to exist; letter C near the center of the top boundary of the figure (at age ~6 b.y.) indicates when the nuclear reactions have proceeded sufficiently to reduce the H abundance to zero in the core. **At this stage of evolution, the material closest to the core of the star consist almost entirely of helium alone, as shown by the label He in the white area near the lower right-hand corner of the left-hand panel. The helium core persists at later times, as shown by the He label in the white areas of the middle panel and the right-hand panel. The letter D, also along the top boundary, and situated near the right-hand edge of the left-most panel** in Fig. 1 (at age 7.1 b.y.) indicates when the convective envelope has grown to contain 10% of $M_*$. The time axis is expanded after 7.2 b.y. **(in the middle and right-hand panels)** in order to demonstrate the effect that is of **primary interest to us in the present paper**. At times 7.2 - 7.46 b.y., the H-burning shell moves *outwards* in the mass coordinate from roughly $0.15M_*$ to $0.29M_*$. At the same time, the convective envelope (denoted by the presence of many interlocking curly lines) moves *inward*, with its lower boundary eventually penetrating down to a maximum depth when it contains a mass coordinate M(c) = 0.76$M_*$ (at age 7.38 b.y.). **Of primary significance in the context of the formation of the "bump", we note that a major consequence of the penetration of the convective envelope (with its initial abundance of hydrogen which remained unburned in the cooler near-surface layers of the star) inward to any radial location r is to homogenize the local gas at location r with the initial H abundance: this homogenization has the effect of erasing any molecular weight gradients that had built up at earlier times due to H-burning in the core**. At times *t* > 7.38 b.y., the base of the convective envelope starts to retreat back towards the surface. While the envelope is retreating, the H-burning shell **(lying deeper than the base of the convective envelope)** continues its own outward movement. At some point in time (7.41 b.y. in this case), the H-shell reaches mass coordinate M(c) and suddenly encounters gas which (because convection had at an earlier time "passed through") contains a lot of unburnt H. With sudden access to richer H-fuel, the enhanced energy generation causes the shell to expand, cooling it slightly. This cooling causes the overall luminosity of the star to *decrease temporarily* by several percent. After a finite time (depending on the magnitude of the discontinuity in H abundance), equilibrium is restored and the star continues its *upward* motion along the RGB. Because the star spent extra time in the neighborhood of the temporary decrease in luminosity, when one calculates the luminosity distribution function for a cluster of stars, there is a local enhancement in the number of stars at that luminosity. **This enhancement in the luminosity distribution is called the RGB "bump".**

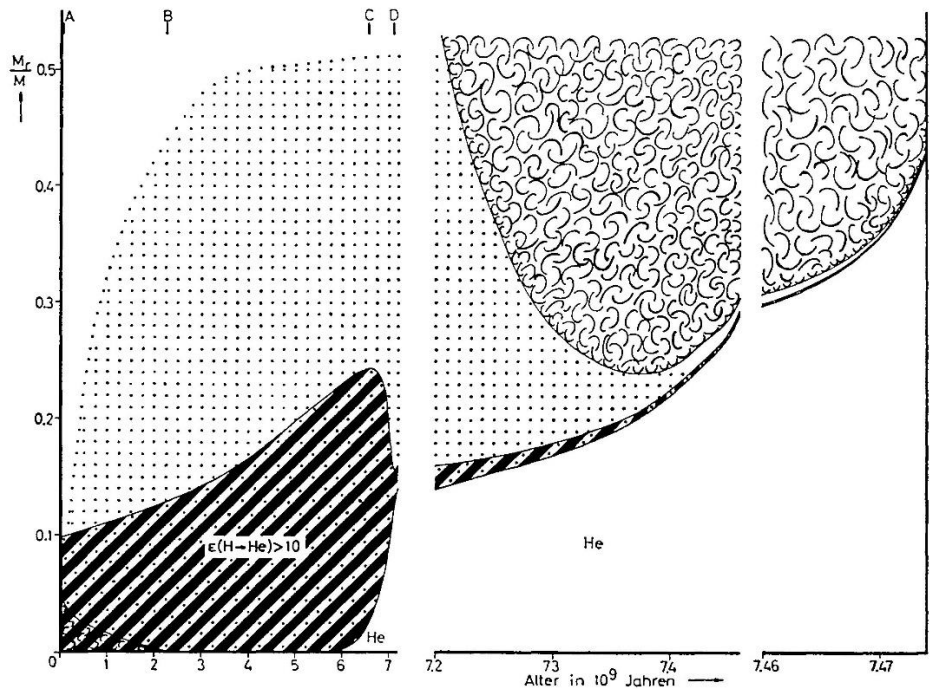

Figure 1. Illustrating the origin of the "bump" (from Thomas 1967).

A detailed study of the properties of the "bump" for clusters of stars having a variety of ages and metallicities has been provided by (among others) Cassisi & Salaris (1997). Over the observed range of metallicities which are relevant for galactic globular clusters, the absolute visual magnitude of the bump is found to range from as bright as -0.3 to as faint as +1.3. Thus, the stars at the bump have luminosities of 100 $L_\odot$ (for the most metal poor stars) or 25 $L_\odot$ (for the most metal rich stars). With effective temperatures at the bump of order 4800 K (Cassisi & Salaris 1997), **these numerical values of luminosity require that** bump stars have radii of 7-15 $R_\odot$ depending on metallicity. In our discussion below, we shall adopt a radius of 10 $R_\odot$ as being typical for a bump star.

### 2.3. Observational effects at the bump

Attention has recently been paid to the RGB bump by Bharat Kumar et al. (2015): these authors infer that "during the short span of bump evolution, giants probably undergo some internal changes, which are perhaps the cause of mass-loss … events". Apparently, Bharat Kumar et al. (2015) were not aware that Mullan & MacDonald (2003) had earlier proposed precisely the same empirical correlation between the bump and the onset of cool massive winds in K giants. Moreover, a detailed physical mechanism for the onset of bump mass loss had also been presented (MacDonald & Mullan 2003: see discussion in Section 3 below).

As a further aspect of bump evolution, **Bharat Kumar** et al. also find that there is evidence for excess lithium abundances once the star has evolved through the bump. In fact, prior to the bump, the lithium abundance becomes increasingly depleted as a star evolves towards the bump from below (Hill et al. 2019: see their Fig. 7). But once the stars have evolved though the bump, certain stars **(but not all)** are observed to have unexpectedly **higher lithium abundance by amounts which may be as large as 1-2 dex**. Apparently, something happens to some RGB stars during bump evolution that can cause an alteration in the lithium abundance. Hill et al. (2019) discuss several possible explanations for the high Li-abundance that they discovered in one particular giant star in the Sculptor dwarf galaxy without coming to a definitive conclusion as regards the origin of the high Li.

As well as alterations in lithium abundance at the bump, certain K giants that show far-IR excess radiation (i.e. evidence for warm dust, presumably due to mass loss) also begin to appear at or within the bump region. In fact, early data (De La Reza et al. 1996, 1997) suggest that there might be a physical connection between lithium enrichment and mass loss in K giants. In the present context, it is especially noteworthy that in a sample of seven K giants that exhibit both Li-richness as well as far-IR excess *all* are found to be situated in the HR diagram in the bump region (Bharat Kumar et al. 2015). However, Bharat Kumar et al. (2015) suggest that although bump evolution may be "perhaps the cause of mass-loss and Li-enhancement events", the sample is small enough that there is not yet enough evidence to prove that these properties are correlated in a statistically significant way. **However, in a discussion of enhanced mass loss in bump stars, MacDonald and**

**Mullan (2003) had already proposed an explanation of enhanced lithium in such stars: we will return to this topic in Section 3.6 below after we have explained the properties of shear instability.**

Now that we have identified observational signatures **which suggest** that something interesting is occurring at the bump, we turn to a particular examination of the physical characteristics of the material inside a **rotating** star when it evolves up along the RGB and reaches the bump.

3. SHEAR INSTABILITY IN ROTATING RED GIANTS

In the present paper, we are especially interested in a process that can occur under certain conditions in rotating stars, namely, shear instability. This is distinct from convective instability: the latter sets in if the local gradient of *temperature* becomes steeper than a critical value (the Schwarzschild criterion). In contrast, shear instability requires that the local gradient of *velocity* becomes steeper than a critical value (the Richardson criterion). In both types of instability, the onset of instability causes the local gas to start moving in a more or less turbulent manner by tapping into the local source of free energy. Dramatic simulations of shear instability can be seen online in articles dealing with Kelvin-Helmholtz flows: turbulent flows are clearly generated as the instability develops. (We note that the Earth's atmosphere contains both types of instability: convective instability causes turbulence in a thundercloud, while shear instability gives rise to "clear air turbulence" in regions of strong wind shear.)

For many years, at least since the paper by Parker (1955), the ability of turbulent flows arising from *convection* have been used to model how dynamos can be powered in rotating astrophysical objects. Here, we suggest that, in a region of the interior of a star where convection is absent, this does not necessarily mean that a dynamo will not operate: on the contrary, we suggest that the flows arising from *shear instability* may also be relied upon to generate magnetic fields in **rotating** stars.

## 3.1. Modeling of differentially rotating stars: shear instability

MacDonald & Mullan (2003) modelled the effects of differential rotation (**in the radial coordinate)** on stellar structure using an assumption of "shellular rotation" (Zahn 1992), in which the angular velocity depends only on depth. Starting from a state of rigid body rotation, angular momentum can be transferred in the radial coordinate via the effects of turbulent transport. As time proceeds, and the star evolves from a large pre-main sequence object to a smaller main sequence object, the star in general speeds up in order to conserve overall angular momentum. Then as the star evolves on the main sequence, the surface rotation rate slows down. However, in the interior of the star, **the process of angular momentum transfer has the effect that the angular velocity** $\Omega$ tends to increase with time. This rotational evolution is what leads to differential rotation in the radial direction as the star evolves. The rotational speed of the star at any particular radial location is not necessarily equal to the speed at a **slightly different radial** location (although the outer convective envelope is in general in solid body rotation). **If, at certain evolutionary phases, the radial gradient of Ω in a region of the stellar interior rises to a critically steep value, this can give rise to onset of shear instability in that region**.

The specific criteria used by MacDonald & Mullan (2003) to determine whether local conditions in the stellar model are stable to shear are

$$Ch > Bh > \frac{D^2}{4},$$

where

$$D = r\frac{d\Omega}{dr},$$

$$C = \frac{1}{\Gamma_1}\frac{d\ln p}{dr} - \frac{d\ln \rho}{dr},$$

and

$$B = -\sum_i \left(\frac{\partial \ln \rho}{\partial \ln X_i}\bigg|_{p,T} \frac{d\ln X_i}{dr}\right).$$

If either of the criteria is violated, shear instability occurs. The criterion

$$Ch > \frac{D^2}{4},$$

is the generalization of the Richardson criterion for Kelvin-Helmholtz instability in plane parallel geometry (Miles 1961; Howard 1961) to shellular rotation (MacDonald 1983; Fujimoto 1987; Hanawa 1987). Flows that are stabilized by density/entropy stratification can be destabilized by thermal diffusion. The criterion

$$Bh > \frac{D^2}{4},$$

expresses the stabilizing effects of chemical composition gradients on shear flow that would be unstable in a homogeneous medium in the presence of thermal diffusion (Zahn 1974; MacDonald 1983). MacDonald & Mullan (2003) generalized mixing length theory for convection to include shear by incorporating these results into the linear stability analysis of Kato (1966).

Of particular importance in the present context is the following: MacDonald & Mullan (2003) found that shear instability occurs as the star evolves through the RGB "bump". At this phase of evolution, it was found that an outer region of shear instability (ORSI) arises **in a particular region of the star**, specifically in the **region** lying between the H-burning shell and the convective envelope. **In this paper, we are especially interested in examining how the existence of shear instability in the ORSI might ultimately contribute to driving mass loss from such a star.**

### 3.2. Shear instability: a critical length-scale

An important property of the shear instability in stars has to do with the rate at which the instability grows. MacDonald & Mullan (2003: see their Section 5.3) have emphasized that, in the context of rotating RGB bump stars in which a radial gradient of angular velocity is present, a critical length scale $L_c$ exists with the following important property: in structures which are smaller than $L_c$, shear instability can grow rapidly (on dynamical time-scales), whereas in structures which are larger than $L_c$, shear instability develops much more slowly (on secular time-scales). To illustrate how large the difference can be between dynamical and secular growth rates, we note that, in one particular example illustrated by MacDonald & Mullan (2003: see especially their Figure 1), the dynamical growth rate is found to be ≥$10^7$ times the secular growth rate. **Such a large difference in growth rates makes it important for us to identify the elements of gas inside the ORSI which will be subject to the fastest instability, i.e. those with sizes smaller than $L_c$ .**

The reason for the existence of a critical length scale can be summarized briefly as follows: although the presence of shear certainly provides free energy to drive instability, this driving can be **counteracted (in part or entirely)** by stabilizing processes due to radial gradients of molecular weight $(\nabla_\mu)$ and/or entropy $(\nabla_s)$. If such stabilizing gradients are present, the shear instability will develop slowly, and will not lead to efficient dynamo activity. But if the stabilizing gradients are absent, then shear instability grows **dynamically,** at least $10^7$ times more rapidly, thereby leading to efficient dynamo generation of magnetic fields.

### 3.3. Mass loss and the "bump": empirical data

Prior to the theoretical work of MacDonald & Mullan (2003), Mullan & MacDonald (2003) had used IUE data to draw attention to an empirical correlation between the bump and the **onset** of cool massive winds in a sample of 15 cool giant stars with a range of metallicities: one of the stars in that sample (Arcturus) will be of particular interest in the present paper in view of the very steep velocity profile of its wind (see Section 5.4 below). The great value of high-resolution IUE data in the context of winds from cool stars was stressed by Stencel & Mullan (1980) when they quantified the asymmetries which are observed in the emission cores of the MgII h and k lines. When stars **were** plotted in an HR diagram, asymmetries indicating outflow were found to appear in stars that lie above and to the right of a locus that Stencel & Mullan referred to as a "velocity dividing line" (VDL). The asymmetries of the emission cores **were** found to be correlated with the wavelength shift at the core of the MgII k line over a range of wavelength shifts up to 0.5 Å (Stencel et al. 1980). Such shifts indicate the presence of outward radial velocities of tens of km s$^{-1}$ at the level in the atmosphere where the core of the MgII h and k lines are formed. Thus, whatever is giving rise to the wind in these stars, the wind has already been accelerated to speeds of 10's of km s$^{-1}$ at locations in the atmosphere where the emission cores of MgII h and k are being formed.

Prior to the work of Stencel & Mullan (1980), Linsky & Haisch (1979) had used IUE data to report on the presence of a "temperature dividing line" (TDL) in the HR diagram such that stars above and to the right of the TDL contained no lines detectable by IUE from plasma having temperatures of 10$^5$ K or hotter. Such stars are referred to as "non-coronal" stars. On the other hand, stars to the left and below the TDL are referred to as "coronal" stars, because IUE successfully detected emission from lines that are emitted by plasma at temperatures above 10$^5$ K. The VDL discovered by Stencel & Mullan (1980) was found to be situated in the HR diagram at a location which is close to (or overlaps with) the TDL.

### 3.4. Mass loss and the "bump": qualitative discussion

Given the empirical correlation between "bump" and onset of mass loss **reported by Mullan and MacDonald (2003),** the question arises: might the onset of shear instability in the ORSI of a bump star have a physical connection with the onset of mass loss? A *qualitative* possibility for such a connection was postulated by MacDonald & Mullan (2003) as follows: in the ORSI, dynamical instability leads to rapid development of eddies with a range of length-scales. There exists a critical length-scale, $L_c$, that separates eddies having distinctly different stability properties. For eddies which are larger than $L_c$, thermal diffusion time-scales are so long that the temperature fluctuations continue in existence for some time before diffusing away: as a result, entropy gradients persist long enough to regulate these large eddies, which as a result develop quite slowly. In contrast, for eddies which are smaller than $L_c$, entropy gradients are wiped out on short time scales, removing the regulatory aspects of entropy gradients. In the absence of such regulation, the small eddies are free to develop rapidly on a dynamical time-scale. **This leads is to propose the following hypothesis which is central to the idea proposed in the present paper**: the dynamo that operates in the ORSI preferentially creates magnetic flux *on short length-scales.* If this is correct, **then the following corollary is a reasonable expectation**: a star on the RGB which lies *above* the bump, should contain on its surface a population of short loops *which have no counterpart in a star lying below the bump.*

## 3.5. Mass loss and the "bump": quantitative discussion

The key question concerning RGB bump stars can now be framed as follows: are the physical conditions in any region of the star such that the growth of shear instability can be rapid? To answer this, we note that MacDonald & Mullan (2003) reported on the presence of two distinct regions of shear instability (RSI) in these stars. One is the *inner* RSI (labelled IRSI) which lies on the inner edge of the H-burning shell. The second is the *outer* RSI (labelled ORSI) which occupies the region between the outer edge of the H-burning shell and the base of the convective envelope. In evolutionary terms, the ORSI does not make its appearance until the convective envelope has started to retreat towards the stellar surface, i.e. until the star has passed through the RGB bump.

It is important to consider the difference between the IRSI and ORSI as regards their effectiveness in promoting instability (and thereby generating magnetic fields). The material in the IRSI has two important features that have an impact on the growth rates of instability: (a) the gas is convectively stable, i.e. the value of $\nabla_s$ is significant, and (b) there is a significant radial gradient of the composition $\nabla_\mu$. Both of these gradients reinforce each other in the IRSI to reduce the growth rate of shear instability: the instability grows only at the slow (secular) rate. **In such conditions, we expect that** magnetic field generation is **relatively ineffective.**

On the other hand, in the ORSI, because the convective envelope has already "swept through" the local material, the gas has been stirred up so as to reduce $\nabla_\mu$ to zero. Thus, one of the stabilizing factors that was present in IRSI is no longer relevant. What about the second stabilizing factor: $\nabla_s$? This factor can also become irrelevant in the ORSI because local temperature fluctuations in eddies with length scales $L$ can be actively erased by a diffusive process on a thermal time-scale $\tau_{th} = L^2/\kappa_{th}$, where $\kappa_{th}$ is the local thermal diffusivity. Because of the presence of $L$ in the expression for $\tau_{th}$, the value of $\tau_{th}$ becomes progressively smaller as we consider smaller eddies. At a critical length scale $L_c$, the value of $\tau_{th}$ falls to values that are so short that they **become shorter** than the dynamical time-scale $\tau_{dyn} = H/c_s$.

(Here, $H$ is the local scale height, and $c_s$ is the local sound speed.) As a result of this, eddies which are smaller than $L_c$ lose their thermal identity on time-scales which are so short that the temperature fluctuations disappear on time-scales which are smaller than $\tau_{dyn}$. In such eddies, therefore, the second stabilizing factor ($\nabla_s$) also becomes inoperative. This has the effect that in ORSI eddies with length scales $\leq L_c$, the shear instability can grow on the (fast) dynamical time-scale, thereby causing vigorous turbulence. **In such conditions, we consider it plausible that the fast-growing (dynamical) instability in the ORSI can generate magnetic fields effectively.**

What typical values might be expected for the critical length scale $L_c$ in RGB bump stars? By equating $\tau_{th}$ with $\tau_{dyn}$, we see that $L_c = \sqrt{(H\kappa_{th}/c_s)}$. Inserting $H = R_gT/\mu g$ and $c_s = \sqrt{(\gamma R_gT/\mu)}$, where $R_g = 8.2 \times 10^7$ c.g.s. is the gas constant, we find $L_c = (\kappa_{th}/g)^{0.5}(R_gT/\mu\gamma)^{0.25}$ where all quantities are to be evaluated in the ORSI. Since the values of $\mu$ and $\gamma$ are of order unity, the value of $(R_g/\mu\gamma)^{0.25}$ is of order $10^2$. To estimate the value of $g$, MacDonald & Mullan (2003) have listed parameters for various evolutionary stages of a 1 $M_\odot$ star: for their stages 7 and 8 (surrounding the bump), we find that at the base of the convective envelope (i.e. close to the ORSI), the radial coordinate is 3 – 4 $\times 10^{10}$ cm, and the mass coordinate is 0.24 – 0.29 $M_\odot$. This leads to a local value of $g \approx 2 \times 10^4$ cm s$^{-2}$. Therefore, $L_c$ is of order $\kappa_{th}^{0.5}T^{0.25}$. Furthermore, since the ORSI lies close to the H-burning shell, the value of $T$ must be of order $10^7$ K, leading to $L_c \approx 50\ \kappa_{th}^{0.5}$. The value of $\kappa_{th}$ depends on the physical process which is at work in the ORSI to reduce the thermal fluctuations in the ORSI.

Since the time-scale on which the shear instability develops is an evolutionary time-scale of order $t_e \approx 10^8$ yr $\approx 3 \times 10^{15}$ s, and the relevant length scale $H$ (the pressure scale height), is comparable to the local radial coordinate (3 – 4 $\times 10^{10}$ cm), the turbulent diffusivity $\kappa_d \approx H^2/t_e$ must be of order $3 \times 10^5$ cm$^2$ s$^{-1}$ in the ORSI. This is significantly smaller than the radiative thermal diffusivity $\kappa_r$, which **(in the models computed by MacDonald and Mullan [2003]) is found to be** of order $3 \times 10^9$ cm$^2$ s$^{-1}$. Thus radiative thermal diffusivity is the dominant process, and we must set $\kappa_{th} \approx 3 \times 10^9$ cm$^2$ s$^{-1}$. **Using this**, we derive an estimate of $L_c \approx 3 \times 10^6$ cm. At the ORSI radial location (3 – 4 $\times 10^{10}$ cm), the value of $L_c$ is smaller than the radial coordinate by some 4 orders of magnitude. **In view of this small value of $L_c$, we see that there is room to accommodate** as many as $N$(ORSI) $\approx 4\pi \times 10^8 \approx 10^9$ eddies on the ORSI surface which are small enough to serve as dynamical-scale dynamos.

The magnetic fields generated in these eddies will lead to buoyancy forces which drive the magnetic eddies upwards toward the surface of the star. Although some of these eddies will undoubtedly experience dissipation (by diffusion or by reconnection) before reaching the surface, a finite number $N_b$ of these eddies may be able to survive all the way to the surface. The essential assumption of the present paper can be stated as follows: we suggest that those magnetic flux tubes from the ORSI that survive long enough to reach the surface of the star can serve as $N_b$ discrete sources of plasmoids emerging into the stellar wind. **We shall estimate a numerical value for $N_b$ below.**

How large will the eddies which survive from the ORSI to the surface be when they reach the surface? An eddy which starts off at the ORSI as an object of size $L_c$ = 3 × 10$^6$ cm would have an initial diffusive lifetime $\tau_{th}$ = $L_c^2/\kappa_{th}$ of order 3 × 10$^3$ seconds in a medium with $\kappa_{th}$ = 3 × 10$^9$ cm$^2$ s$^{-1}$. However, a magnetic flux tube emerging from such a structure and rising up via buoyancy forces through the star, will expand its size L above the initial value that we have estimated for $L_c$. In order to estimate the eddy sizes on the surface, we note that if the mass of the rising flux tube is conserved in the course of the rise, $L$ will increase as $\rho^{-1/3}$. The models indicate that the gas density in the ORSI (where $L = L_c$) is of order $\rho_O$ = 0.08 g cm$^{-3}$, while the density in the photosphere **of the star at the same point in evolution** is of order $\rho_s$ = 5 x 10$^{-8}$ gm cm$^{-3}$. As a result, the size $L_{surf}$ is expected to exceed $L_c$ by a factor of order 100-200. That is, $L_{surf}$ is expected to be of order 3000-6000 km. Such structures are certainly small compared to the radius of a "bump" star: we are therefore justified in referring to a population of "small loops" on the surface. **It is of interest to compare the values we have derived for $L_{surf}$** with the length scales that have already been considered for plasmoids in the solar corona: 250-1000 km (e.g. Pneuman & Cargill 1985; Mullan 1990). Thus, we are proposing the presence of plasmoids in RGB "bump" stars which are **somewhat** larger in size with those which have already been used in models of fast (plasmoid-driven) wind from the Sun.

An important difference between a "bump" star and the Sun concerns the mass associated with a single plasmoid. In the Sun, the typical mass $M_n$ of a nanoflare-generated plasmoid is of order 10$^8$ g (Mullan 1990). Energetic arguments suggest that the number of nanoflares occurring per second $N$(nf) is of order $10^5$ (Mullan 1990): this number is essentially equal to $N_n/T_n$, where $N_n$ is the number of potential nanoflare sites on the surface at any instant, and $T_n$ is the time interval required to braid the local magnetic field, forcing it to become unstable, thereby initiating a nanoflare. Estimates of the time interval required to drive a coronal loop unstable to flaring can be made, based on convective jostling of field lines (Mullan & Paudel 2018): the times **in the case of the Sun** are estimated to be between $T_n \approx 10^{4.5-5.8}$ s. Thus, $N_n$ may be of order $10^{10}$. If the nanoflare process was perfectly efficient at ejecting mass, the rate of ejection of mass $dM/dt$ from the Sun should be of order $N(nf)$ times $M_n \approx 10^{13}$ g s$^{-1}$. However, the mean solar mass loss rate is measured to be $10^{12}$ g s$^{-1}$: therefore the efficiency $\varphi$ of plasmoid mass ejection **cannot be any larger than** 10%, i.e. $dM/dt \approx \varphi M_n N_n/T_n$. In a "bump" star, the eddy mass ($\sim L_c^3 \rho$) is $M_b \approx 2 \times 10^{18}$ g: the mass loss rate due to ejection of such plasmoids will be $(dM/dt)_b \approx \varphi M_b N_b/T_b$ where $N_b$ is the number of small loops present on the surface of the "bump" star at any instant, and $T_b$ is the time interval to drive one of those small loops unstable. Now, in Section 2.1 above, we noted that the empirical data suggest that the average mass loss rate on the RGB may be as large as $(dM/dt)_b \approx 10^{-8}$ M$_\odot$ yr$^{-1}$, i.e. roughly $10^6$ times larger than the solar value ($2 \times 10^{-14}$ M$_\odot$ yr$^{-1}$ : see Linsky & Wood 1996). In terms of grams per second, $(dM/dt)_b \approx 10^{18}$ g s$^{-1}$. This requires that $\varphi N_b/T_b$ should have a numerical value of order 1 s$^{-1}$, i.e. $N_b/T_b \approx 10$ (assuming the same value of $\varphi$ as in the Sun). How might the flare-generation time-scales $T_b$ in a "bump" star compare with those in the Sun? In the presence of convective granules with size $\lambda$ and convective speed $v$, the value of $T_b$ is estimated to be of order $L^2/\lambda v$ for a loop with footpoint size $L$ (Mullan & Paudel 2018). For a "bump" star, the gravity is 10-100 times smaller than in the Sun: the temperature is also smaller than solar. As a result, $\lambda \approx T/g$ will exceed the solar value by a value of at least 10. On the other hand, we have estimated above that the length scale of a "small loop" is 3000 km: as a result, the footpoint of such a loop may be of order $L = 1000$ km, i.e. 1 order of magnitude smaller than the smallest value assumed for loops in the Sun (Mullan & Paudel 2018). Convective velocities in a "bump" star are expected to be comparable to solar values: avoidance of shock formation requires that the flows remain subsonic. Therefore, in a "bump" star, the combination of a decrease in $L^2$ by perhaps $10^2$, and an increase in $\lambda$ by perhaps 10, we estimated that $T_b \approx L^2/\lambda$ will be smaller than the smallest value in the Sun ($10^{4.5}$ s) by a factor of order $10^{2-3}$. This leads to $T_b \approx 10^{1.5-2.5}$ s, and therefore $N_b \approx 300$-$3000$ on the surface of a "bump" star at any instant. Given that some $10^9$ small eddies can exist on the ORSI surface (see discussion above), we are requiring that only 1 in $10^{5.5-6.5}$ of those small eddies **manage** to survive the passage from the ORSI to the surface of the star. If this requirement can be fulfilled, then the plasmoid model of mass loss in a "bump" star leads to a mass loss rate for such a star that is comparable to the observed magnitude ($10^{-8}$ $M_\odot$ yr$^{-1}$).

In view of the relatively small number ($N_b \approx$ 300-3000) of small loops which are present on the surface of the "bump" star at any given instant, it seems likely that the mass loss rate of a plasmoid-driven wind from a "bump" star will be subject to temporal fluctuations at the fractional level of $1/\sqrt{N_b}$, i.e. at the 2-6% level. In this regard, Crowley et al. (2009) have pointed out that "There is mounting evidence" that winds from cool giants exhibit mass loss which is "inhomogeneous in both space and time". Quantitative statements to this effect can be found in a study by Mullan et al. (1998) in which the optical thickness of the **winds from two cool giants** was found to change by factors of 2-6 from one epoch to another, although admittedly the stars in that study were not close to the "bump", but considerably farther into the regime of "non-coronal" stars.

### 3.6. A population of small loops above the bump

Why might the presence of a population of small loops be significant in the context of mass loss **from cool giants**? In the context of the VDL proposed by Stencel & Mullan (1980), Rosner et al. (1995) suggested that the onset of mass loss in cool giants arose because of "a change in the size scale of the magnetic loops in the atmosphere": in particular, when loops become *small*, Rosner et al. suggested that this leads to enhanced mass loss rates. However, Rosner et al. did not specify any particular mechanism that would lead to smaller loops, nor did they explain why such a mechanism (whatever it was) should occur at a particular location in the HR diagram. In particular, Rosner et al. made no reference to the RGB bump.

In view of the discovery of shear instability (specifically, the ORSI) in association with the RGB bump, MacDonald & Mullan (2003) suggested that the onset of massive loss above the bump might be attributable to the emergence of a population of small loops on the surface.

MacDonald & Mullan (2003) also pointed out a corollary that bears on the lithium excesses mentioned above (Section 2.3). If magnetic flux tubes are in fact created by the ORSI, then *the upward* buoyancy of the **rising flux tubes has the following important accompanying effect: the rising tubes impede** the *downward* diffusion of material from the surface layers of the star. With the hindering of downward gas motions, lithium depletion (which requires the surface gas to reach deep layers where the Li can be subject to nuclear burning) **is rendered** less efficient than if there were no rising magnetic fields.

In the present paper, we wish to examine in more detail what happens when a multitude of small magnetic loops emerge at the surface of a cool giant star. The examination will be conducted in the context of a particular "plasmoid" model proposed by Pneuman (1983) in order to account for fast wind flows from the Sun. When the paper by MacDonald & Mullan (2003) appeared, there was not sufficient information available **in the literature** concerning magnetic field strengths and **upper** atmospheric temperatures **in cool giants** to allow the plasmoid model to be applied quantitatively to the problem of interest to us here. **However, in the course of** the past few years, **some of the relevant** information has become available. It is therefore timely to attempt **at least a proof-of-concept by means of a** quantitative application of the plasmoid model of mass loss from stars at the bump.

## 4. MODELS OF MASS LOSS FROM COOL GIANTS

Before we **examine** the plasmoid model **in the context of cool giant stars**, it is worthwhile to describe some of the efforts that have been made to model the winds **from such** stars: these will provide guidance for choosing certain physical parameters in our models.

### 4.1. Ordered fields

Hartmann & MacGregor (1980) proposed a model for accelerating steady-state winds from cool giants using Alfven waves propagating outward along radial field lines. The approach was based on the method of Jacques (1977). However, in order to avoid **driving the winds so hard that the** terminal velocities **become** too large, the Alfven waves must dissipate within a few stellar radii of the surface. The dissipation **of the Alfven waves** should lead to extended warm chromospheres that would be an integral part of a turbulent region where the wind is accelerating outward. Comparison between theory and empirical line profiles in Hubble data suggests that the model is not altogether successful in accounting for certain aspects of the observed line profiles: e.g. predicted Doppler widths are larger than the observed turbulence, the wind and the chromosphere may not constitute a single unit (Eaton 2008), and sharp acceleration close to the stellar surface is observed in some stars while others show evidence of a slower (or delayed) acceleration (Crowley et al. 2009).

In passing, we note that Hartmann & MacGregor (1980) placed the base of their wind models in gas with densities of $10^{11} - 10^{12}$ cm$^{-3}$. In the model to be presented in this paper, we shall find that our estimates of the density at the base of the wind we propose for a bump star **overlaps** the range **of densities** used by Hartmann and MacGregor.

### 4.2. Chaotic fields: clumpy wind

An alternative magnetically-driven wind mechanism was proposed by Eaton (2008). Rather than relying on the globally organized Alfven waves used by Hartmann & MacGregor (1980), Eaton suggested that chaotic fields might emerge from the star, "diffusing through the gas into space to drag the gas along with it and away from the star". In this model, the wind from a cool giant is viewed as being clumped, i.e. bifurcated into dense clumps separated by an unspecified interclump medium. Moreover, rather than the magnetic field remaining rooted in the star (to serve as a conduit for Alfven waves), magnetic flux would be lost from the star along with the gas in the wind. **Eaton (2008) parametrized the** clumping of the wind **in terms of** a volumetric filling factor $f$, related to a Clumping Factor (CF = $1/f$). In order to fit data for the binary 31 Cyg, Eaton **found** that CF is required to rise to a maximum value of 10 - 15 at **radial locations of** 1.5 – 2 R$_*$, and then decrease to values of CF = 5 - 10 at radial locations out to 4 – 5 R$_*$. Such values of CF correspond to volumetric filling factors $f$ = 0.06 - 0.2. Eaton estimated that the surface fields in his model would have to be stronger than 25 - 35 G in order to overcome gravity. In **Eaton's** model, the mass-loss rate of a star would vary with time depending on how the rate of magnetic field emergence at the surface varies in different stages of evolution.

Harper (2010) examined the Eaton model in the context of continuum radio fluxes for a different binary system (ζ Aur) belonging to the same class of eclipsing binaries as the star modelled by Eaton (31 Cyg). Harper **found** that the CF values for the wind of ζ Aur rise to a peak value of ~10 close to the star's surface (1.1 R∗), and then occupy a plateau with CF of ~5 throughout the subsonic region of the wind. The radial profiles of CF seem**ed** quite similar for 31 Cyg and ζ Aur, but Harper notes that "in the region of interest, the CF for ζ Aur are a factor of 2 smaller than for 31 Cyg". In order to optimize the fit between model and radio data at wavelengths of 2.0, 3.5, and 6.2 cm, Harper suggest**ed** that the best compromise might be provided by CF = 2.25 (i.e. $f$ = 0.44): such a large filling factor, in Harper's opinion, should not be referred to as "a highly clumped magnetized wind". Nevertheless, Harper (2010) agrees with Eaton (2008) in the following statement: "clumping, or rather structure, still may be an important aspect of cool star wind models".

In the present paper, we accept this statement and proceed to examine a quantitative model for a clumpy wind emerging from a star at the bump stage of RGB evolution.

## 5. PNEUMAN'S MODEL: ACCELERATION OF SOLAR WIND CONTAINING DIAMAGNETIC PLASMOIDS

### 5.1. The solar case

Pneuman (1983) noted that enormous amounts of magnetic flux loops are emerging through the surface of the Sun ON SMALL SCALES all the time. Magnetic reconnection **in the upper photosphere/chromosphere/corona** allows this flux to be dissipated and/or expelled on a time scale equal to the emergence scale. **As a result of r**econnection, **the loop can become pinched off,** with the upper part becoming a self-contained magnetic plasmoid (see Figure 2). In the presence of an ambient magnetic field $B_a(r)$ which systematically weakens as the radial distance $r$ increases (see Figure 3), **a plasmoid with a self-contained internal magnetic field (which does not penetrate into the ambient field) is accompanied by a surface current j. As a result of the Lorentz force j X $B_a$,** the plasmoid experiences a net radially *outward* force which is proportional to the product of the plasmoid mass, the ambient temperature, and the radial gradient of the magnetic pressure $p_B$, where $p_B = B_a^2/8\pi$. The outward force that leads to driving of the individual plasmoids causes the ambient gas to be dragged outwards also. Colloquially, this process has been described as the "melon-seed mechanism" (Schlüter 1957).

Pneuman's model does NOT require a large gas pressure to drive mass loss: the driving requires only that the overall magnetic field of the star falls off with increasing distance from the star. Therefore, this model can produce winds in stars even where the gas pressure is low. **This aspect of the model suggests that the melon-seed mechanism could be useful for driving a wind from a cool giant.**

Pneuman defines the parameter $R$ = ratio of mass flux of diamagnetic material to the total mass flux. Thus, the Pneuman constant $R$ can be regarded as a sort of "mass filling factor" describing the fraction of the overall mass flux which is carried by the plasmnoids. In Pneuman's formalism, **he makes the assumption that** the numerical value of $R$ **remains** constant throughout the wind. In this case, the location of the critical point occurs at a radial location $r_c = GM_*/[2a^2(1+3R)]$, where $a$ is the isothermal sound speed of the ambient gas. In the limit $R = 0$, i.e. in the absence of plasmoids, **we recover the well-known radial location of the Parker (1963) sonic point in an isothermal corona, i.e. $r_c = GM_*/2a^2$**. In the presence of finite $R$, the critical point is moved inward (i.e. the acceleration is steeper) **with a reduction in the radial location of the critical point** by a maximum factor of 4 if $R = 1$.

It does not seem necessary to consider only the case in which $R$ remains constant at all radial locations. In view of the finite possibility that material may be able to exchange from time to time between plasmoids and the ambient medium (due, e.g. to reconnection, or diffusion), we cannot exclude the possibility that $R$ might vary as a function of radial position. A particular example in which $R$ was allowed to vary with radial location has been presented by Mullan (1990): in that example, the filling factor of plasmoids increased from a very small value close to the surface to essentially 100% at a certain radial location. In the present paper, rather than introducing more free parameters, we retain the Pneuman simplification $R$ = constant.

An interesting feature of the melon-seed mechanism is that, **when the internal gas pressure in the plasmoid is included,** the magnetic force $F$ acting on a diamagnetic plasmoid is **found to be** proportional to the mass $M$ of the plasmoid times the radial gradient of the logarithm of the ambient magnetic pressure (see eq. (19) in Pneuman [1983]). As a result, the melon-seed mechanism leads to an acceleration of the plasmoid ($a = F/M$) which is independent of the mass. As we have seen in Section 3.5 above, plasmoid masses in "bump" stars are expected to be orders of magnitude larger than the masses of plasmoids emerging from nanoflares in the Sun. **Nevertheless,** despite this difference, the acceleration of a plasmoid in both cases will be the same provided that the radial gradient of $\log(B_a)$ is the same (where $B_a$ is the ambient magnetic field through which the plasmoid flows).

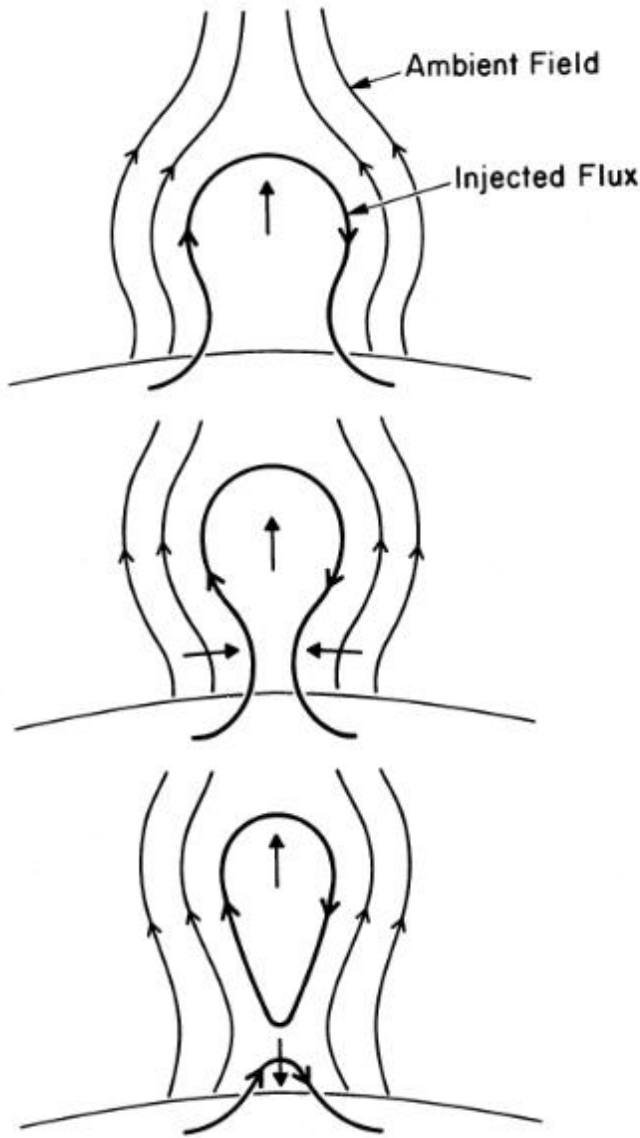

Figure 2. Schematic of a reconnection event as illustrated by Pneuman (1983). Emergent flux pushes aside the ambient field as the new flux rises. Restoring forces are set up in the displaced ambient field that pinch off the loop via magnetic reconnection. Reconnection leads to the formation of an isolated bubble (or plasmoid) which continues to rise. © AAS: used with permission

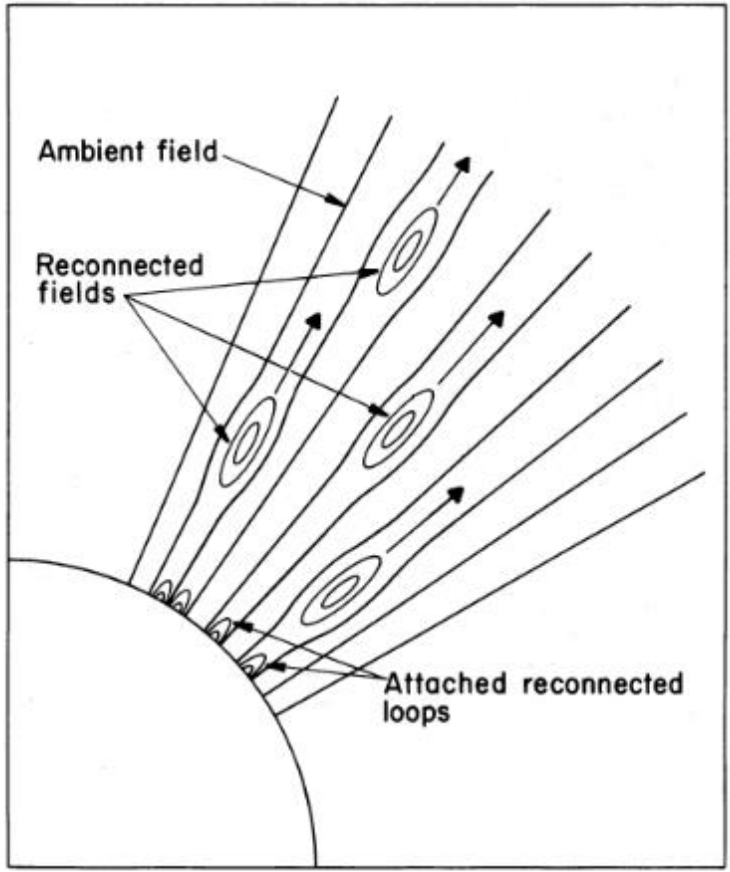

Figure 3. Topology of ambient and ejected fields in the Pneuman (1983) model. Each reconnection ejects a closed magnetic plasmoid in the upper regions, and leaves an attached closed loop at the surface of the star. In the presence of the ambient magnetic field, which diverges outward, each plasmoid is subject to a systematic outward force. © AAS: used with permission.

### 5.2. Application to a cool giant

Inspection of Figure 3 suggests an application of the Pneuman model to the case of a cool giant at the bump. In particular, we **draw the reader's attention to** the multitude of small-scale loops on the surface of the "star" in Figure 3. We suggest that these are analogous to the "small loops" proposed on heuristic grounds by Rosner et al. (1995) for a "non-coronal star" above and to the right of the VDL. We also suggest that the small loops on the stellar surface in Fig. 3 **owe their origin in a "bump" star** to the "small loops" arising from the ORSI discovered by MacDonald & Mullan (2003) as occurring in a rotating star when it evolves through the bump.

In order to apply the Pneuman model to a star other than the Sun, there are two important physical parameters that we need to specify in order to quantify the plasmoid-driven wind.

### 5.2.1. Temperature of the atmosphere

In order to determine the motion of the plasmoids as they accelerate out through the stellar atmosphere/wind, the temperature of the ambient medium must be specified **(see eq. (19) of Pneuman [1983])**. At first sight, when we combine the results of Mullan & MacDonald (2003) with those of Linsky & Haisch (1979), it would seem that stars above the bump should fall into the category of "non-coronal" stars. In the earliest days of stellar X-ray astronomy, such an expectation might appear to have been supported by the lack of detectable X-ray emission from such stars even when the most sensitive detectors on the ROSAT X-ray satellite were used for deep pointings (Ayres et al. 1991). However, within the past year, using the greater sensitivity of the X-ray detectors on the Chandra satellite, the prototype "non-coronal star" (Arcturus) has been discovered to be a faint but statistically significant source of X-rays in the 0.2 - 2 keV range of energies (Ayres 2018). The Chandra data suggest that the ratio of X-ray luminosity to bolometric luminosity for Arcturus is $L_x/L_{bol} = 5 \times 10^{-11}$. This ratio is almost 4 orders of magnitude smaller than the equivalent ratio for the long-term average Sun, itself already known to be a star which has close to the lowest level of activity among solar-like stars in the solar neighborhood. Thus, after a search that has lasted for 3-4 decades, Arcturus is finally established to be truly a source of X-rays, albeit a faint source.

The most important property that we would like to know about the X-ray source in Arcturus is the coronal temperature. The best way to derive this quantity would be to have an X-ray spectrum. Unfortunately, the Arcturus X-rays are so faint that no spectrum is yet available: **as a result**, no precise estimate of the coronal temperature can be currently obtained. However, Ayres (2018) has argued that, using a particular model (O'Gorman et al. 2013) of the wind density for Arcturus, one can estimate the hydrogen column density of the outflow to be of order $8 \times 10^{19}$ cm$^{-2}$. Then using this wind column density as input to a web-based simulator, as well as a guess for the possible coronal temperatures, one can calculate the Energy Conversion factor (ECF) which enables conversion from an observed counting rate (CR, in units of **Chandra** counts per kilosec) of photons in the Chandra data into an apparent (un-absorbed) X-ray flux at Earth. For coronal temperatures in the range 2.5 - 10 MK, Ayres (2018) finds ECF = $8.1 \pm 0.3 \times 10^{-15}$ erg cm$^{-2}$ s$^{-1}$ (cnt ks$^{-1}$)$^{-1}$. Moreover, if the coronal temperature is lower, i.e. in the range 1 - 2 MK, the ECF changes by only about 10%. Ayres concludes that the systematic error in the X-ray flux which would reach Earth is only about 10% even if the coronal temperature is allowed to range from as low as 1 MK to as high as 10 MK. In other words, it is not inconsistent with the available data to guess that the corona in Arcturus has a temperature of 1 MK. **In support of this estimate**, it is difficult to imagine that photons with energies of at least 0.2 keV (such as those detected by Chandra from Arcturus) could be emitted by gas that is significantly cooler than 1 MK. In view of that, we will assume that Arcturus contains gas with temperatures of order 1 MK.

In some cool giants (e.g. Betelgeuse), evidence exists that the winds are inhomogeneous in space and time (Crowley et al. 2009), with regions of hotter plasma interspersed among cooler material. In view of this, it is possible that the coronal material (at T = 1 MK) **in the atmosphere/wind of Arcturus might be interspersed among cooler material. Since the X-ray emission from Arcturus is observed to be weak, it is possible that the filling factor of the coronal material might be small. If this were to be the case, a referee** has pointed out that the model proposed in the present paper would not be valid. However, the plasmoids we consider in this paper have a well-defined characteristic that is relevant in this context, **namely,** the plasmoids undergo expansion as they rise upwards into lower density material. Quantitative modeling of how much expansion actually occurs depends on the properties of the plasmoids, especially if they are magnetically dominated or pressure dominated (Pneuman & Cargill 1985; Mullan 1990). These models show that as a plasmoid rises from the solar surface to a height of one solar radius above the surface, the radius of the plasmoid can expand in a radial field by factors of up to 10 – 30. And if the field expands super-radially, the plasmoid may grow in radius by up to a factor of 100 as it propagates from the solar surface to a height of one solar radius above the surface. These expansions of plasmoids in radius as they move outward have the effect that even if the coverage factor or coronal material might be small near the stellar surface, the volume filling factor of the plasmoids grows with increasing distance from the star. In fact, in one of the plasmoid models calculated by Mullan (1990) **for the case of the Sun**, the following conclusion emerged: "by the time the plasmoids .... have risen to altitudes of about one solar radius above the surface, they may fill essentially the entire available volume". Thus, even if the surface coverage of hot coronal gas in Arcturus turns out to be "patchy", this deficiency is **offset** in the context of wind outflow because of the significant expansion which plasmoids undergo during their outflow.

In what follows, when we apply the Pneuman model to stars on the RGB, we will assume that the ambient medium has a temperature of 1 MK. The corresponding isothermal sound speed $a = \sqrt{R_g T / \mu}$ in an ionized gas with $\mu$ = 0.6 is 118 km s$^{-1}$.

### 5.2.2. Magnetic field strength

The strength of the magnetic field on the surface of a single K giant must be specified if we are to apply Pneuman's model to cool giants at the bump. Measurements of magnetic field strengths in stars need to be interpreted carefully. If the measurements are based solely on circular polarization data (i.e. Stokes V), then the data refer mainly to the longitudinal component of the field (i.e. along the line of sight): **in such data**, there is significant cancellation of oppositely directed fields which are present **in different regions on** the visible hemisphere of the star. **On the other hand, m**easurements of Stokes I offer greater sensitivity to localized areas of strong field on the surface of the star. The mean field strengths obtained from V and I data are such that the ratio $<B_V>/<B_I>$ for M dwarfs turns out to be at most 15%, and in other cases no more than a few percent (Morin et al 2010). **As a result**, the "true" value of the mean field strength on the surface of a star is larger by a factor ψ of at least 6 (and possibly as large as ψ ≈ 20) than the field strength suggested by Stokes V data. In the case of the Sun, the referee has pointed out that even more extreme differences can exist between the values of $B_V$ and $B_I$ : in spatially resolved data, $B_I$ can be of order 1500 G (with a small filling factor), whereas when the unresolved Sun is observed in Stokes V, the field $B_v$ is found to be about 1 G: **in this admittedly extreme case**, the factor ψ might be as large as $10^3$.

In the present context, we note that Auriere et al. (2015) observed a sample of 48 G-K giants and obtained **statistically significant usable** Stokes V data in 29 of their targets: for these 29, they reported the maximum unsigned longitudinal magnetic field strength $|B_l|_{max}$. In the case of 8 stars where they were able to apply the techniques of Zeeman Doppler Imaging, Auriere et al. found that the surface-averaged magnetic field strength was found to compare well with $|B_l|_{max}$ suggesting that the field strengths are **more or less** reliable. The magnitudes of the field strengths were found to range from ~100 G to 0.25 G. In view of the cancellation of flux inherent in Stokes V data, these field results are lower limits to the fields which may exist on the surface**s of G-K giants.**

The stars in the Auriere et al (2015) sample which are of most interest to us, i.e. the stars which lie closest to the bump, are Pollux (K0 III), Arcturus (K1.5 III) and Aldebaran (K5 III). For these stars, Auriere et al. reported $|B_l|_{max}$ values of 0.7, 0.34, and 0.25 G respectively. Two of these three stars have the weakest fields that were discovered by Auriere et al. If we apply the correction factor ψ of "at least 6" mentioned above, a better estimate of the mean field strength on the surface of a "bump" star might be at least 1.5-4 G. If ψ ≈ 20 (see discussion of the work on stellar data by Morin et al [2010] above), a "bump" star might have "true" fields of order 10-20 G. And in the most extreme case, if we were to apply the value of ψ = $10^3$ suggested by solar data, a "bump" star might have localized fields (with small filling factors) of almost 1 kG. Therefore, in constructing a plasmoid wind model in the present paper, we shall consider <B> values in the range from a few G up to 1 kG for purposes of making estimates. **It must be admitted that the range of field strengths which we consider here is so wide that we can make no claim to obtaining a detailed model of the wind: we re-iterate that our goal in this paper is to determine if a proof-of-concept yields plausible results.**

### 5.3. Plasmoid wind: overall mass-loss rate

Consider a star near the RGB bump with radius $R_* = 10\ R_\odot$ (see Section 2.2). On the surface of this star, we envision that there exist multiple small loops undergoing magnetic reconnection as in Fig. 3 above. In the simplest model of reconnection (Sweet-Parker), the material which emerges from the reconnection process moves at the Alfven speed $v_A = B/\sqrt{4\pi\rho}$ which is characteristic of the fields and densities of the material in the ambient medium (Parker 1957). With the field lines of the small loops arranged as in Fig. 3, the material emerging upward from a reconnection site lying above any of the small loops **(i.e. lying in the space between the top of a small loop and the bottom of its overlying plasmoid)** are expected to flow in the upward direction at speed $v_A$. Suppose that the reconnection sites are sufficiently numerous as to cover a fraction $f$ of the surface area of the star, i.e. $f$ is an areal filling factor for the mass outflow. (We shall assume that the areal filling factor $f$ and Pneuman's mass filling factor $R$ are interchangeable.) The mass-loss rate from the entire surface of the star is expected to be

$$-\frac{dM}{dt} = 4\pi R_*^2 f v_A N m_H, \qquad (1)$$

where $N$ is the number density of the gas containing the reconnecting fields, and $m_H$ is the mass of a hydrogen atom. In principle, the density which is to be used in the definition of $v_A$ refers to the ionized component of the material. However, in the presence of frequent ion-neutral collisions, the neutral atoms are also dragged along by the ions that are forced into motion by a passing Alfven wave (e.g. Mullan 1971). Therefore, the relevant density to be used in the Alfven speed is the total number density (mainly neutrals) in the medium.

As mentioned above (Section 2.1), we require a mass loss rate for a star at the bump to be of order $10^{18}$ g s$^{-1}$. Substituting $R_* = 10$ R$_\odot = 7 \times 10^{11}$ cm, we find that the observed mass loss rate can be supplied provided that the product $Bf\sqrt{N}$ at the base of the wind (where reconnection is occurring) has the value $5 \times 10^5$ in cgs units. In the most efficient case, where $f \sim 1$ (we shall label this case, Pneu = 1 in Fig. 4 below), this requires $N$ to have a value of order $(3/B^2) \times 10^{11}$ cm$^{-3}$. If $B = 2$ G is inserted (see Section 5.2.2), we find that $N$ is of order $10^{11}$ cm$^{-3}$. In a less efficient case, where $f = $ (say) 0.1, we find that $N$ must have a value of order $(3/B^2) \times 10^{13}$ cm$^{-3}$. If we insert $B = 2$ G, we find $N \approx 10^{13}$ cm$^{-3}$, but if $B$ is as large as 20 G, then we find $N$ is of order $10^{11}$ cm$^{-3}$. And if an even smaller $f$ value is inserted, $f = 0.01$, we find that with $B = 20$ G, **the result is** $N \approx 10^{13}$ cm$^{-3}$.

Are such densities plausible for the wind of a cool giant star? When Hartmann & MacGregor (1980) were computing their models of Alfven wave driven steady-state winds, we note that they choose number densities at the base of the wind in the range between $10^{11}$ and $10^{12}$ cm$^{-3}$. Thus, our estimates of N turn out to overlap with those which were adopted at the base of the completely independent model of Hartmann and MacGregor (1980). Even if we go to the extreme case mentioned above, and insert $B = 10^3$ G, we can still recover $N = 10^{12}$ cm$^{-3}$ provided that the filling factor f is ≈ 6 x $10^{-4}$. How does such a filling factor compare with what we have estimated above concerning the sizes and number of "small loops" on the surface of a "bump" star? With sizes D of up to 6000 km, and numbers of loops N(L) up to 3000 on the surface (see Section 3.5), the fractional area coverage of the loops on the surface of a star with radius $R_* = 10R_\odot$ is expected to be f = $D^2N(L)/4\pi R_*^2$ ≈ 2 x $10^{-4}$. Although this is not quite as large as the above value of f derived above in the extreme case, nevertheless, in view of the many contributing uncertainties to the various parameters in the model, it is worth noting that we have "come close" to consistency in terms of our model of multiple "small loops" on the surface of a "bump" star.

We conclude that, with allowance for uncertainties in some of the parameters, the physical conditions are such that **a plausible number** of discrete reconnection sites are capable of supplying the mass flux required to explain the wind emerging at a mass loss rate of $10^{-8}$ M$_\odot$ yr$^{-1}$ from a cool giant star at the RGB bump.

### 5.4. Radial profile of wind speed

The equation of motion derived by Pneuman (1983) for the outflow speed (his eq. 24) is identical to the equation derived by Parker (1963) for the hydrodynamic outflow of an isothermal corona, except for a modification to one of the terms. The term $2a^2/x$ in Parker's equation (where $a$ is the sound speed and $x = r/r_c$ is the radial distance scaled to the radius at the critical point in the wind, i.e. the point where the flow speed equals the sound speed) is replaced in Pneuman's equation by the term $2a^2(1+3R)/x$. The critical point in Parker's equation occurs at $r_c$(Pa) = $GM_*/[2a^2]$, while in Pneuman's equation, the critical point occurs closer to the star, at $r_c$(Pn) = $GM_*/[2a^2(1+3R)]$.

In the Appendix, the equation of motion which is used in this paper is described in detail. One aspect of the equation which differs from that of Parker or Pneuman is that although the coronal temperature is held constant over most of the wind, there exists a small region close to the star (within a few tenths of a stellar radius) where the sound speed is assumed to approach zero: this ensures that when we consider giant stars, the critical point will not lie **beneath the stellar surface**.

Using eq. (10) in the Appendix, we have calculated the radial profile of the outflow speed in the solar wind using different values of the Pneuman filling factor *R*. In Figure 4, we present results for the velocity profile in cases where we insert 4 values for the Pneuman constant *R*: 0.0 (i.e. the Parker model), 0.1, 0.2, and 1.0. The position of the Parker critical point for a corona with T = 1 MK (i.e. with isothermal sound speed a = 118 km s$^{-1}$) lies at $r_c$(Pa)= 6.9 R$_\odot$. In the case of the velocity profiles plotted in Fig. 4, with *R* = 0.1, 0.2, and 1.0 the Pneuman critical points are predicted to lie at $r_c$(Pn)= 5.3, 4.3, and 1.7 R$_\odot$ respectively.

In the case *R* = 0.0 (i.e. the Parker solution), the slow acceleration of the wind solution in Fig. 4 is evident: at a radial distance of 3 stellar radii, the speed has still reached a value of no more than ~50 km s$^{-1}$. But in the case where *R* = 0.2, the wind speed reaches 100 km s$^{-1}$ within a distance of ~4 stellar radii. Even faster acceleration occurs in the case *R* = 1 (the maximum permissible value): the wind speed reaches 150 km s$^{-1}$ within a radial distance of order 2 stellar radii, i.e. the wind accelerates from zero to more than 100 km s$^{-1}$ within an altitude of no more than 1 stellar radius above the surface.

Although the results in Figure 4 pertain strictly to the wind from a star which has the radius of the Sun, **it is worthwhile to compare some features which occur in Fig. 4** with certain features of the velocity profiles which have been reported for **cool** giant stars (e.g. Crowley et al. 2009, their Figure 1). In particular, the fast acceleration in the case *R* = 1 in Fig. 4 is reminiscent of the steep radial profile that has been reported for Arcturus, where the wind speed reaches terminal velocity within less than one stellar radius of the surface. Moreover, the values we have chosen for the Pneuman (mass) "filling factors" for the cases *R* = 0.2 and 0.1 overlap with the (volumetric) filling factors that have been extracted empirically by Eaton (2008) and by Harper (2010) for 31 Cyg and ζ Aur, i.e. f = 0.06-0.2.

These smaller filling factors, with the shallower velocity profiles as shown in Fig. 4 (where sonic speeds are not reached until the radial location is of order 5 stellar radii), are also consistent with the shallower velocity profiles reported for the primary star in two ζ Aur binaries, where distances of 7 stellar radii or more are required to reach terminal speed (Crowley et al, their Fig. 1).

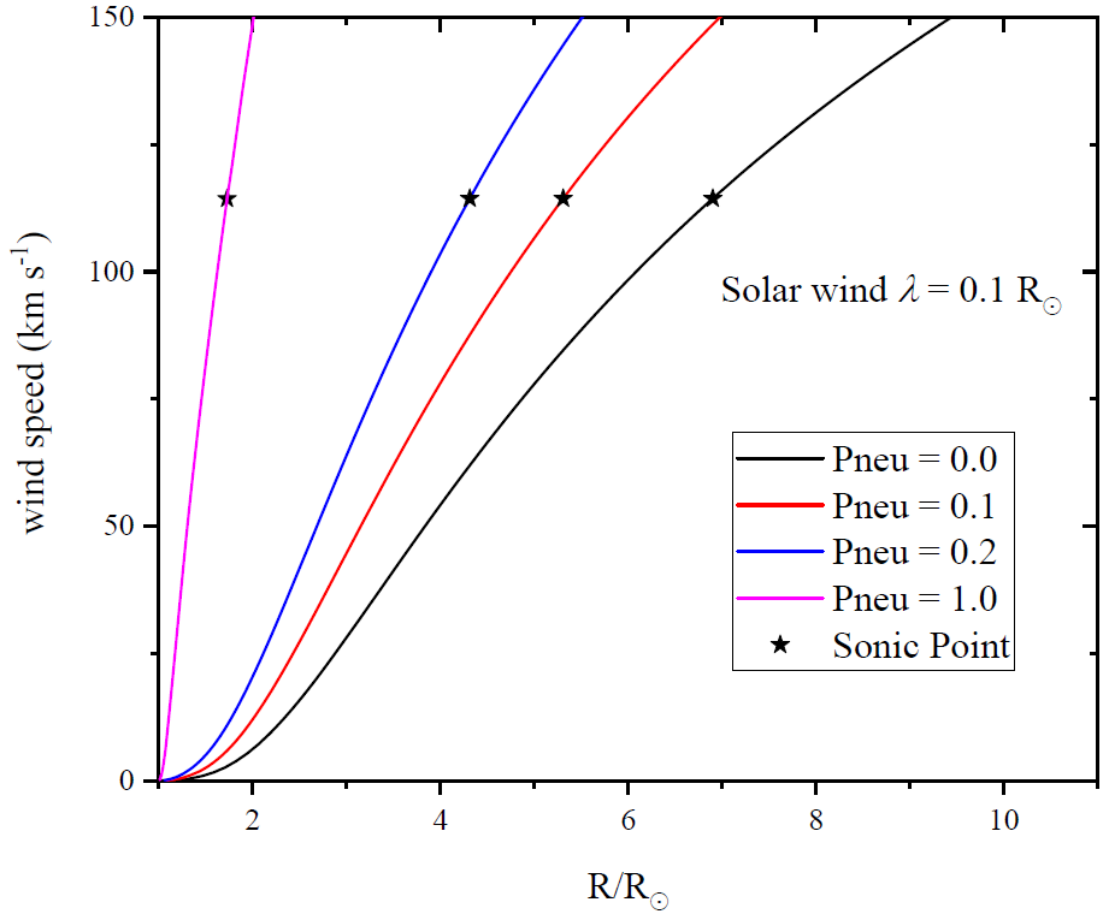

Figure 4. Radial profiles of wind speed from the Sun in cases where the Pneuman filling factor $R$ takes on numerical values of 0.0, 0.1, and 0.2, and 1.0. For the sake of clarity, in this figure, Pneuman's $R$ is **written** as Pneu to avoid confusion with the radius parameter. The radial coordinate is normalized to the solar radius.

In a third category of wind velocity profiles **in cool giant stars**, Crowley et al. (2009) have reported (for SY Mus and EG And) that the wind velocity remains small until the radial location exceeds 2-3 stellar radii, and then accelerates quickly. However, both SY Mus and EG And are symbiotic systems consisting of a red giant (RG) and a white dwarf (WD) separated by a few RG radii. This arrangement leads to RG wind structures that, at distances of a few RG radii, depart significantly from spherical symmetry (Shagatova et al. 2016). We do not consider these systems further in this paper.

6. DISCUSSION AND CONCLUSIONS

Cool **giant stars with spectral types** K through mid M are known to lose mass at a rate that can be as large as several times $10^{-8}$ M$_\odot$ yr$^{-1}$, i.e. larger than the solar mass loss rate by as much as 6 orders of magnitude. The existence of these massive winds, as well as mechanisms to drive them, has been the subject of discussion for decades. However, apart from ruling out certain processes (e.g. pulsations, radiative driving of dust), a recent summary of the situation contains the not atypical statement: "the mechanisms that drive mass-loss [from these stars]…are very poorly understood" (Harper 2018).

In the present paper, we have attempted to address this deficiency by considering a model developed by Pneuman (1983) to account for high speed solar wind: the model is based on the empirical result that magnetic flux is observed to be continually emerging upwards through the solar surface. As the magnetic flux rises through the solar atmosphere, local conditions may **give rise to the onset of magnetic reconnection. As a result, some of the magnetic flux experiences a process of "pinching off"**, thereby creating an isolated plasmoid. In the presence of an ambient (global) magnetic field that steadily diminishes in strength with increasing distance from the Sun, such plasmoids are subject to a radially outward Lorentz force (known colloquially as the "melon-seed mechanism"). Pneuman (1983) has quantified the amount of coronal acceleration experienced as a result of the presence of plasmoids which contribute a fraction $0 < R < 1$ to the overall mass outflow.

The possibility that a connection might exist between the solar magnetic processes considered by Pneuman (1983) and those in certain cool giants arose in a qualitative sense some years ago **when we were considering** a model of shear instability in the course of **the evolution of a rotating cool giant star** (MacDonald & Mullan 2003). That discussion led to the proposal that in the course of evolution, especially at the **evolutionary** feature known as the "bump" on the RGB (Thomas 1967; Iben 1968), the magnetic fields on the surface of a cool giant would suddenly become dominated by small-scale fields. The reason for the emergence of small-scale fields at the "bump" has to do with a competition between dynamical time-scales and thermal diffusion time-scales in a region of a rotating star where the gas is subject to shear instability: on small length scales, dynamical processes **dominate over diffusive processes** and cause the small-scale fields to grow more rapidly than those on larger length scales (see Section 3.5 above). As pointed out by Mullan & MacDonald (2003), the location in the HR diagram where cool massive winds are first observed to set in occurs in close association with the "bump". This suggested that the emergence of small-scale fields associated with shear instability might contribute to driving mass loss in cool giants.

The fact that Pneuman (1983) had demonstrated that the emergence of small-scale magnetic fields into the solar atmosphere can provide an extra **(Lorentz)** force which contributes to rapid wind acceleration suggested that it might be worthwhile to see if an analogous process is relevant for the small-scale fields emerging when a cool giant star evolves through the bump. However at the time when the MacDonald & Mullan (2003) paper appeared, there was no empirical information available as regards either the magnetic field strengths or the coronal temperatures in the stars we are interested in. In the past few years, this lack of information has been remedied to a certain extent (Auriere et al. 2015; Ayres 2018).

With access to the new information, **the present paper represents a return to examine if the Pneuman formalism can be applied to cool giants in the vicinity of the RGB bump. The information which is currently available, although not yet sufficient to allow us to develop a detailed model, nevertheless allows us to examine a proof-of-concept. Using permissible ranges of various parameters, the estimates we have obtained here suggest that the empirically determined rate of mass loss on the RGB can be accounted for. Moreover, we find that the radial profile of wind speed can range from rather gradual to quite steep, depending on the mass filling factors of the plasmoids in the wind: the** latter result could explain empirical studies that indicate that certain cool giants have steep accelerations while other cool giants have more gradual accelerations.

Acknowledgements

We thank the referee Jeffrey L. Linsky for a constructive and helpful review of an earlier draft of this paper. The work of JM is supported in part by the Delaware Space Grant program.

**APPENDIX:** Stellar wind equations

Here we derive the equations that we use to model a stellar wind which includes the mechanism proposed by Pneuman (1983). We assume that the wind is in a steady state so that the momentum and mass conservations equations for a star of mass $M$ and mass outflow rate $F$ can be written as

$$\rho v \frac{dv}{dr} = -\frac{dp}{dr} - \rho \frac{GM}{r^2}, \qquad (2)$$

and

$$F = 4\pi r^2 \rho v, \qquad (3)$$

where $v$, $p$, and $\rho$ are the wind speed, pressure, and density, respectively. Writing

$$p = \rho a^2, \qquad (4)$$

where $a$ is the isothermal sound speed (which is assumed to be a function of $r$), equation (2) becomes

$$v\frac{dv}{dr} = -\frac{1}{\rho}\frac{d\rho}{dr}a^2 - \frac{da^2}{dr} - \frac{GM}{r^2}. \qquad (5)$$

Using equation (3) to eliminate $\rho$, equation (5) becomes

$$\left(v - \frac{a^2}{v}\right)\frac{dv}{dr} = \frac{2a^2}{r} - \frac{da^2}{dr} - \frac{GM}{r^2}. \qquad (6)$$

From Pneuman (1983), inclusion of the 'melon-seed' mechanism modifies the first term on the right hand side of equation (6) so that it becomes

$$\left(v - \frac{a^2}{v}\right)\frac{dv}{dr} = \frac{2a^2}{r}(1+3R) - \frac{da^2}{dr} - \frac{GM}{r^2}. \qquad (7)$$

where $R$ is ratio of the mass flux of diamagnetic gas to the total mass flux. Making the substitution

$$v^2 = ya^2, \qquad (8)$$

equation (7) becomes

$$\left(1 - \frac{1}{y}\right)\frac{dy}{dr} = \frac{4}{r}(1+3R) - (1+y)\frac{d\ln a^2}{dr} - 4\frac{a_\infty^2}{a^2}\frac{R_p}{r^2}, \qquad (9)$$

where $a_\infty$ is the sound speed at infinity and

$$R_p = \frac{GM}{2a_\infty^2}. \qquad (10)$$

The solution of equation (9) has a critical sonic point at a position given by the solution of

$$4\frac{a_\infty^2}{a^2}\frac{R_p}{r^2} = \frac{4}{r}(1+3R) - 2\frac{d\ln a^2}{dr}. \qquad (11)$$

The gradient of *y* at the critical point can be found, by differentiating equation (9), to be a solution of

$$\left(\frac{dy}{dr}\right)^2 + \frac{d\ln a^2}{dr}\frac{dy}{dr} = \frac{4}{r^2} - 2\frac{d^2\ln a^2}{dr^2} - 2\left(\frac{d\ln a^2}{dr}\right)^2 + \frac{12R}{r}\left(\frac{d\ln a^2}{dr} + \frac{1}{r}\right). \tag{12}$$

To construct figure 4, we use a specific functional form for the sound speed:

$$a^2 = a_\infty^2\left[1 - \exp\left(-\frac{r - R_*}{\lambda}\right)\right], \tag{13}$$

where $R_*$ is the stellar radius and $\lambda$ is a heating length scale, presumably related to the process(es) which heat(s) the corona. In the results shown in Fig 4, we have taken the heating length scale to be $\lambda = 0.1 R_*$. According to eq. (14), the sound speed in the "corona" is therefore zero at the surface of the star, but the sound speed rises to essentially $a_\infty$ within a few tenths of a stellar radius.